\documentclass[useAMS,usenatbib]{mn2e}
\usepackage{graphicx}

\newcommand\apj{ApJ}
\newcommand\apjs{ApJS}
\newcommand\apjl{ApJL}
\newcommand\araa{ARA\&A}
\newcommand\mnras{MNRAS}
\newcommand\aap{A\&A}
\newcommand\aj{AJ}
\newcommand\pasj{PASJ}
\newcommand\pasp{PASP}

\title[The Discrepancy of MIR Continuum \& PAHs' Features]{
The Discrepancy of Mid-Infrared Continuum and the Features of Polycyclic Aromatic Hydrocarbons for $Spitzer$ \& $Herschel$ SWIRE-field Galaxies
}
\author[Y. N. Zhu and H. Wu]{%
Yi-Nan Zhu$^{1}$\thanks{E-mail: zyn@bao.ac.cn},
Hong Wu$^{1}$\\
$^{1}$Key Laboratory of Optical Astronomy, National Astronomical Observatories, Chinese Academy of Sciences, Beijing 100012, China\\
}

\begin{document}

\date{Accepted.  Received; in original form}


\maketitle

\begin{abstract}
On the basis of the observations of $Spitzer$ and $Herschel$, we present and analyze the correlations among various monochromatic 
infrared (IR) luminosities for star-forming galaxies, selected from two northern SWIRE fields. The 24 \& 70$\mu$m luminosities 
(L[24] \& L[70]), dominated by the continuum of very small grains (VSGs) and warm dust in thermal equilibrium respectively, 
correlate with on-going star formation tightly. The contribution from cool dust excited by evolved stars also increases 
as the wavelength increaseas in the far-infrared (FIR) wavelength range. The spectral features of ionized polycyclic aromatic 
hydrocarbons (PAHs) around rest-frame 8$\mu$m are exited by moderated radiation field related with evolved stars as well, rather 
than by intensive radiation field related with young stars. Even though the carriers of PAHs could be treated as some types 
of VSGs with smaller scale, the radiation condition between PAHs and classic VSGs seems to be significant different. The formulae 
to calculate the Total-infrared luminosity $L_{TIR}$ by using L[8(dust)] \& L[24] are re-scaled, and we find that the L[8(dust)] 
(L[24]) formula may probably underestimate (overestimate) $L_{TIR}$ for the galaxies with remarkable current star formation activity.
\end{abstract}


\section{Introduction}

The dust emission heated by ultraviolet (UV) and optical photons dominates the infrared (IR) output for majority of galaxies. 
From the beginning of the Universe, about half of the bolometric emission of galaxies has been transported to 
Mid-infrared (MIR) to Far-infrared (FIR) \citep{hauser01}. Since the celestial MIR \& FIR photons suffer serious extinction 
when they run into the atmosphere surrounding us, the observations above the absorbers, such as some types of molecules like 
water and carbon dioxide, are essential for IR astronomy. Since the successful launch of $Infrared~Astronomical~Satellite$ 
($IRAS$) in 1983, more advanced space-based infrared telescopes had been launched, such as $Infrared~Space~Observatory$ \citep[$ISO$;][]{kessler96}, 
$Spitzer~Space~Telescope$ \citep{werner04}, $AKARI$ \citep{murakami07}, $Herschel~Space~Observatory$ \citep{pilbratt10}
\footnote{Herschel is an ESA space observatory with science instruments provided by European-led Principal Investigator consortia and with important participation from NASA.}, and $Wide$$-$$field~Infrared~Survey~Explorer$\citep[$WISE$;][]{wright10}.

Dust is considered to be formed in the outer atmosphere of evolved stars, then is ejected out to interstellar 
space by stellar wind or some types of supernovae. So far, many models have been proposed to scale the emission of dust \citep[see, e.g.,][]{mathis89,desert90}, 
and one of them is the silicate-graphite model \citep{mathis77,draine84}. Recently, on the basis of the observations of $Spitzer$, 
a new version of silicate-graphite model has been demonstrated by \citet{draine07b}. This model contains different grain species 
including silicate \& graphite \& PAH (stand for polycyclic aromatic hydrocarbon, \citet{leger84, puget89}), whose abundance 
and size distribution were derived to fit the average IR spectral energy distribution (SED) and reproduce the average extinction 
curve of Milky Way. Since then, this model has been revised by some studies, such as \citet{galliano11}, who used the 
$Spitzer$ and $Herschel$ images of Large Magellanic Cloud (LMC) to restrict the behavior of diverse dust grains.

Without consideration of the dust emission heated by Active Galactic Nuclei (AGN), the FIR luminosity ($>$100$\mu$m) of a 
galaxy is generally dominated by emission of dust in thermal equilibrium with temperature of about 15$-$20K, which will be 
treated as cool dust throughout this paper; while the FIR luminosity ($<$100$\mu$m) is dominated by the dust in thermal equilibrium 
with the temperature of higher than 30K, which is normally refered as warm dust. The MIR continuum ($<$30$\mu$m) comes 
from the emissions of VSGs (stand for very small grains, \citet{cesarsky00}) heated stochastically by single UV photon released by young stars in star formation regions.

MIR continuum is also characterized by some broad emission features \citep{gillett73, willner77}, which are realized to 
be from PAHs. In these years, the understanding on the emissions of PAHs has been prompted based on the observations of 
$ISO$ 
and $Spitzer$, and a tight correlation between the emissions of PAHs and cool dust has been found and studied \citep{haas02, boselli04, bendo06, bendo08, lu08}.  
Low metallicities and the presence of AGNs could lead to the weakening of PAHs emissions \citep{weedman05,siebenmorgen04,verma05,wu07,engelbracht08}.  
In a previous work \citep{zhu08}, on the basis of the $Spitzer$ Wide-area Infrared Extragalactic Survey \citep[SWIRE;][]{lonsdale03}, 
we also found two tight correlations between 24$\mu$m vs. 70$\mu$m luminosities, and 8$\mu$m vs. 160$\mu$m luminosities, respectively. 

Because of the complication and diversity of IR emission, \citet{dale02} employed three MIPS (stand for Multiband Imaging Photometer for $Spitzer$, \citep{rieke04}) bands, 24 \& 70 \& 160$\mu$m 
fluxes to compute total IR luminosity ($L_{TIR}$). Nevertheless, for nowaday IR facilities, the sensitivity and angular resolution 
decrease gradually with increasing wavelength. Hence, in most cases, we could only use MIR photometries, such as 8$\mu$m or 24$\mu$m, 
to estimate $L_{TIR}$. On the basis of the observations of $Spitzer$, \citet{boquien10} demonstrated 
that both the metallicity and the star formation intensity can influence the accuracy of the estimation of $L_{TIR}$ if 
only one or two IR wave band photometries were used. Adding the observations of $Herschel$, \citet{galametz13} recalibrated 
the correlations between $L_{TIR}$ and various monochromatic IR luminosities, then presented the discrepancy of $L_{TIR}$ 
between from SED fitting and from scaling of monochromatic IR luminosities. They also shown that the discrepancy was a function of IR color, while the IR color was sensitive to current star formation. 

In this work, we will add the $Herschel$ photometries to investigate the correlations among various monochromatic IR luminosities. The structure of the paper 
is as follows. We describe the construction of our sample and the estimation of multi-wavelength luminosities in $\S$2. The 
major results on correlation analysis are presented in $\S$3. Some discussions and a summary of this work are given in $\S$4 
and $\S$5, respectively. Throughout this paper, we adopted a $\Lambda$CDM cosmology with $\Omega_{\rm m}=0.3$, $\Omega_{\rm \Lambda}=0.7$
and $H_{\rm 0}=70\,{\rm km \, s^{-1} Mpc^{-1}}$.

\section{Data Reduction}

\subsection{$SDSS$ Data}

Because we need the spectral redshift to calculate the distance and luminosities for IR selected galaxies, firstly we downloaded 
a series of processed catalogs from MPA website \footnote{http://www.mpa-garching.mpg.de/SDSS/DR7/}, then merged 
them together. These catalogs contain $927,552$ sources included in the spectral sample of Data Release seven (DR7) of Sloan 
Digital Sky Survey \citep[SDSS;][]{york00}. Besides raw data, the MPA-JHU team has also derived some physical quantities, 
such as spectral redshift, emission line fluxes, stellar masses \& metallicities. 
The data reductions have been descripted detailed by \citet{kauffmann03, tremonti04, salim07}. It should be noted that 
the galaxies in MPA-JHU catalogs, are not restricted to those have been compiled as main galaxy sample \citep{strauss02}. 
Some extended selected criteria have been adopted by the MPA-JHU team, and the specific description about these selected criteria 
could be found in their website.

\subsection{$Spitzer$ Data}

The $Spitzer$ SWIRE is the largest extragalactic survey program among the six $Spitzer$ cycle-1 Legacy Programs. This survey 
contains six regions, with a total field of $\sim$49 deg$^{2}$. Because of the limitation of the coverage of SDSS and $Herschel$, 
just two northern SWIRE fields, Lockman Hole (center: 14 41 00, +59 25 00) \& ELAIS-N1 (center: 16 11 00, +55 00 00) are used 
in this work. The MIR \& FIR photometric catalogs have been supplied to public by SWIRE team. We downloaded these catalogs 
in IRSA website by using General Catalog Query Engine \footnote{http://irsa.ipac.caltech.edu/applications/Gator/}. These catalogs 
could also be found and downloaded in SWIRE team's website \footnote{http://swire.ipac.caltech.edu/swire/astronomers/data/}. 
For each field of SWIRE, there are three photometric catalogs in IRSA website: four  $Spitzer$ Infrared Array Camera \citep[IRAC;][]{fazio04} 
bands (3.6$\sim$8$\mu$m) and MIPS 24$\mu$m have been merged together; while the other two catalogs contain MIPS 70 and 160$\mu$m 
band photometries respectively. 

We cross-matched the IRAC/24$\mu$m catalog with SDSS catalog with the radius of 3$\arcsec$. After removing the repeated-crossing 
sources, $941$ galaxies were left.  
Before cross-matching with the 70 and 160$\mu$m photometries, we first did statistics of their astrometric precision: the 
mean values of 1$\sigma$ position uncertainty are 1.5 \& 3.5$\arcsec$ for the 70 and 160$\mu$m detected sources, respectively.
Then, we used the $941$ galaxies in the optical-MIR catalog to match the 70 and 160$\mu$m catalogs with the radius of 4.5 
\& 10.5$\arcsec$, the 3$\sigma$ position uncertainties. Finally, there were 253 optical-MIR galaxies that have both 70 \& 
160$\mu$m detectable fluxes. It should be noted that the telescope-limited resolution of $Spitzer$ is 18$\arcsec$ \& 40$\arcsec$ 
at MIPS 70 \& 160$\mu$m, much larger than the cross-matching radius we have used. Inevitably, in order to improve the accuracy 
of our sample, the completeness of the sample has been sacrificed and we must have lost many FIR sources. 

\subsection{$Herschel$ Data}

Herschel Multi-tiered Extragalactic Survey \citep[HERMES;][]{oliver12} is one of the key programs awarded on $Herschel$ Space 
Observatory as a Guaranteed Time survey. HERMES has mapped a set of separated fields in the sky with the Spectral and Photometric 
Imaging REceiver \citep[SPIRE;][]{griffin10} at 250, 350 and 500$\mu$m \citep{levenson10,zemcov13}. Two fields in their second 
Data Release (DR2), Lockman-SWIRE and ELAIS-N1-HerMES, could be matched with the northern $Spitzer$ SWIRE field. We downloaded 
the band-merged xID catalogs released in the end of 2013 (DR2) \footnote{http://hedam.oamp.fr/HerMES/release.php}. The photometries 
in DR2 have been improved significantly \citep{wang13} compared with those in the first Data Release \citep[DR1;][]{smith12,roseboom10}. 
In these band-merged xID catalogs, the fluxes at 350 and 500$\mu$m were extracted on 250$\mu$m positions. In order to extract 
the sources from SPIRE image, a Gaussian shaped PSF with the FWHM set to 18.15, 25.15 and 36.3$\arcsec$ at 250, 350 and 500$\mu$m 
was used by HERMES team. The $\sigma$ of the Gaussian shaped PSF at 250$\mu$m is about 7.8$\arcsec$, which was employed 
by us as the radius to match with the SDSS-Spitzer catalog. After matching, 185 sources were left. If we enlarge the matching 
radius to 23$\arcsec$, 3 times the radius we used above, there are total of 188 $Herschel$ sources left. Hence, for the $Herschel$ 
sources, the completeness of our sample is about 98\%.

\subsection{Sample Selection}

Weak AGN-hosting galaxies are very popular in the local universe \citep{ho97}. AGNs can strengthen MIR continuum, and weaken 
PAHs' emission or even destruct the carriers \citep{wu07}. Therefore, the probable contribution from AGN to MIR \& FIR emissions 
would introduce significant bias on the correlations among those monochromatic IR luminosities. In this study, the traditional 
BPT diagnostic diagram: [NII]/H$\alpha$ versus [OIII]/H$\beta$ \citep{baldwin81, veilleux87} was used to remove AGN and composite 
galaxies (starburst $+$ AGN) \citep{kewley06}. In the $185$ galaxies of our sample, $104$ of them (with the signal-to-noise 
ratios (S/N) of the four emission line fluxes $>$ 3) could be treated as star-forming galaxies without contamination by AGNs, 
and are illustrated in the left panel of Figure 1. The dotted curve in this panel is from \citet{kauffmann03}. We also plot 
the \citet{kewley01}'s criteria as dashed curve, which is usually treated as the upper limit of ratios of emission 
lines in HII regions excited by inside or nearby stars. The objects below the dotted line were classified as star-forming 
galaxies; the others were removed, and they could be classified as narrow-line AGNs or composite galaxies. 

The right panel of Figure 1 is the rest-frame B band absolute magnitude (M$_{\rm B}$) distribution of the $104$ star-forming 
galaxies, and the M$_{\rm B}$ was calculated from the SDSS $g$ and $r$-band magnitudes according to \citet{smith02}. If $-$18 
mag is used as the border to distinguish dwarf galaxies \citep{thuan81}, we find only one of them could be treated as dwarfs.

\begin{figure}
\includegraphics[scale=1]{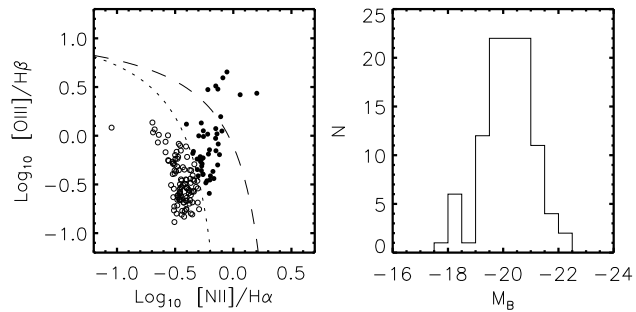}
\caption{The left paned is the BPT diagnostic diagram: [NII]/H$\alpha$ vs [OIII]/H$\beta$. The criteria from \citet{kauffmann03} and \citet{kewley01} are illustrated as dotted and dashed curves, respectively.
The objects below the dotted curve (open circles) are defined as star-forming galaxies;
The right panel is the distribution of rest-frame absolute B-band magnitude for star-forming galaxies.
}
\end{figure}

\subsection{Estimation of Monochromatic Luminosities}

We estimated the monochromatic IR luminosities by using spectral redshift and the MIR \& FIR fluxes provided by above surveys.
For the two $spitzer$ MIR bands, 8 \& 24$\mu$m, the fluxes from kron photometry were adopted as total fluxes; for the two 
$spitzer$ FIR bands, 70 and 160$\mu$m, the PRF (stand for point-response function) photometric fluxes were adopted; for the 
three $Herschel$ FIR bands, 250, 350 and 500$\mu$m, the PSF photometric fluxes were adopted.

Besides the PAHs' features and VSGs' continuum, photosphere of evolved stars could contribute to the output at rest-frame 
8$\mu$m \citep{wu05}, too. Hence, only we could estimate a galaxy's 8$\mu$m dust luminosity after reducing the component of stellar 
continuum. A factor of 0.232 \citep{helou04} was used to scale the star light of 3.6$\mu$m to that of 8$\mu$m, with the assumption 
that the entire IRAC 3.6$\mu$m band emission is from stellar emission, and based on the {\sl Starburst99} synthesis model \citep{leitherer99}, 
assuming solar metallicity and a Salpeter initial mass function (IMF) between 0.1 and 120 $M_{\rm \odot}$. Then, since IRAC 
8$\mu$m band covers the complex PAH emission range, a more accurate SED model is needed to obtain a reliable MIR K-correction. 
Such as in \citet{zhu08}, we adopted a series of SEDs from \citet{huang07}'s two-component model which is a linear combination 
of an old 'early-type' stellar population and a 'late-type' spiral disk population. The MIR color [5.8]-[8.0] was used to 
select the best SED model, and then the K-correction value was obtained. \citet{huang07}'s two-component model was used 
to perform K-correction for the 3.6$\mu$m band as well, which is usually treated as the tracer of stellar mass.

The contribution from stellar continuum to the fluxes of MIPS 24$\mu$m band and the FIR bands can be neglected since the emission 
of the evolved stars' photosphere decreases dramatically with increasing wavelength. A template MIR spectrum of a local 
HII galaxy NGC~3351 (from SINGS, $Spitzer$ Infrared Nearby Galaxies Survey; \citet{kennicutt03}), was convolved by using the 
spectral response curve of MIPS 24$\mu$m band to obtain the K-correction at 24$\mu$m. 

In our sample, all the galaxies' fluxes at 70, 160, 250$\mu$m are larger than 3$\sigma$. Expect one, all the sources' fluxes 
at 350$\mu$m are larger than 3$\sigma$ as well; while the S/N of the rest galaxy's flux at 350$\mu$m is 2.4, and we still 
use it throughout this paper. 61 of the galaxies in our sample have S/N of flux at 500$\mu$m larger than 3$\sigma$.  
For these 61 sources, the K-corrections of five FIR wave band emissions (70, 160, 250, 350, 500$\mu$m) were obtained by fitting 
six bands fluxes (include 24$\mu$m) with a grey body emission and a MIR power-law continuum \citep{casey12}. 
For the other 43 sources, only four FIR wave bands (70, 160, 250, 350$\mu$m) were performed K-correction based on the SED fitting. 

In this work, we use L[a] to refer to the monochromatic luminosity (such as L[24] is the 24$\mu$m luminosity), and from here 
on the convention L[a]=$\nu$L[a] is adopted for IR luminosity measurements. The monochromatic luminosities are in unit of 
solar luminosity, which is 3.83$\times$10$^{33}$ erg/s. In our sample, the mean values of the 1$\sigma$ error of the flux 
from 8 to 500$\mu$m are 0.305\%, 0.714\%, 1.526\%, 2.885\%, 1.901\%, 6.387\%, 15.487\%(the 61 sources with S/N of flux $>$3), 
respectively, corresponding to 0.002, 0.003, 0.007, 0.012, 0.008, 0.026, $0.061$ in dex. 

\section{Results}

\begin{figure*}
\begin{center}
\includegraphics[scale=1]{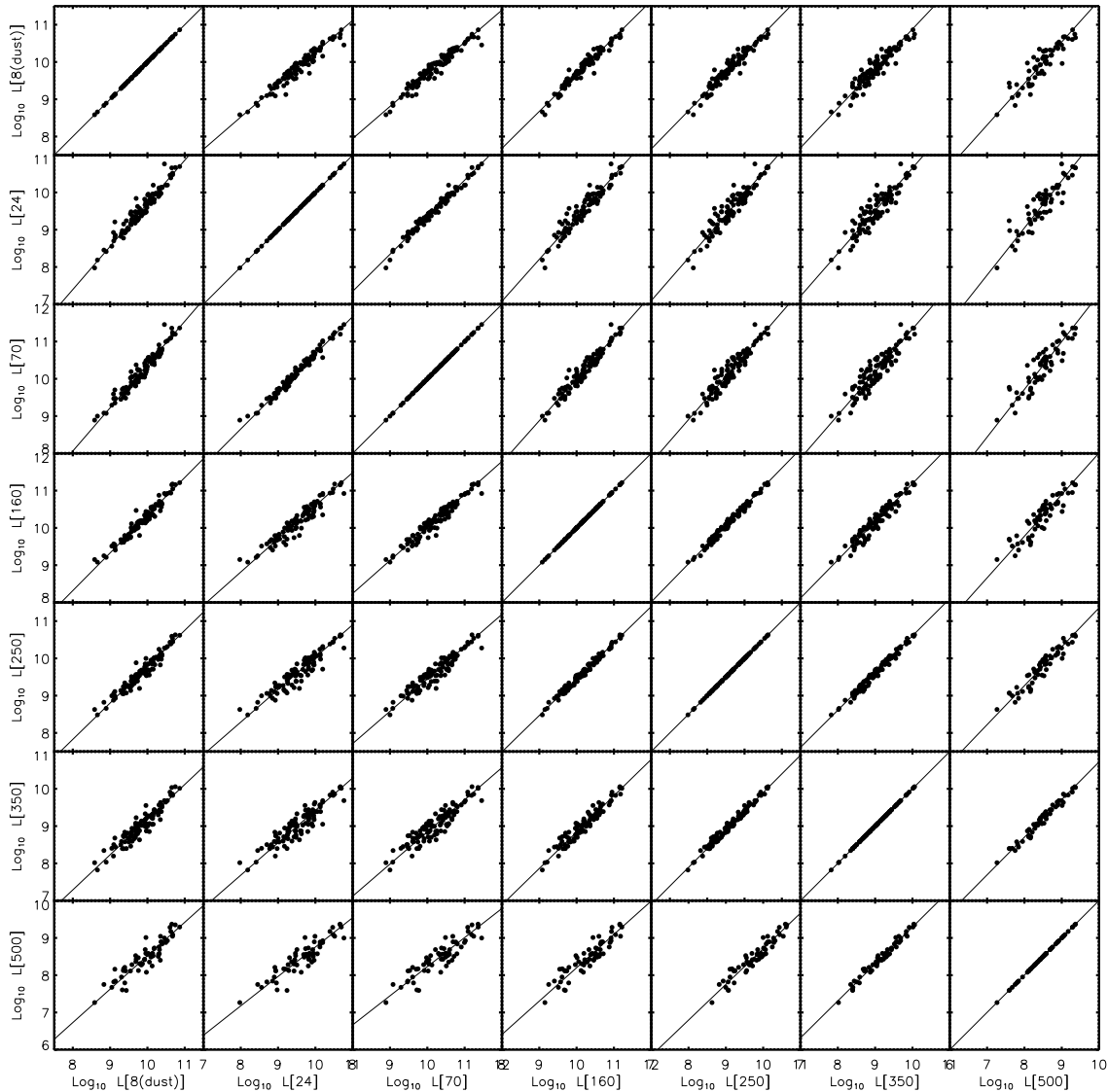}
\caption{Correlations between various monochromatic IR luminosities for star-forming galaxies.}
\end{center}
\end{figure*}

\begin{table*}
\centering
\caption{correlation coefficients between various monochromatic IR luminosities}
\begin{tabular}{llrcccc}
\hline
{$x$} & {$y$} & {$N$}  & {$a$} & {$b$} & {$s$} & {$c$} \\
\hline
{(1)} & {(2)} & {(3)}  & {(4)} & {(5)} & {(6)} & {(7)} \\
\hline
     L[8(dust)]  &   L[24]      &  104 &  -1.112$\pm$0.080 &  1.072$\pm$0.025 &  0.087 &  0.960 \\
     L[8(dust)]  &   L[70]      &  104 &  -1.219$\pm$0.086 &  1.160$\pm$0.027 &  0.092 &  0.971 \\
     L[8(dust)]  &   L[160]     &  104 &   0.228$\pm$0.065 &  1.011$\pm$0.020 &  0.068 &  0.974 \\
     L[8(dust)]  &   L[250]     &  104 &   0.027$\pm$0.073 &  0.975$\pm$0.023 &  0.078 &  0.963 \\
     L[8(dust)]  &   L[350]     &  104 &  -0.257$\pm$0.090 &  0.943$\pm$0.028 &  0.111 &  0.938 \\
     L[8(dust)]  &   L[500]     &   61 &  -0.447$\pm$0.155 &  0.897$\pm$0.049 &  0.139 &  0.898 \\
\hline
    L[24]        &   L[8(dust)] &  104 &   1.038$\pm$0.069 &  0.933$\pm$0.022 &  0.081 &  0.960 \\
    L[24]        &   L[70]      &  104 &   0.016$\pm$0.049 &  1.079$\pm$0.016 &  0.061 &  0.987 \\
    L[24]        &   L[160]     &  104 &   1.199$\pm$0.081 &  0.952$\pm$0.026 &  0.107 &  0.945 \\
    L[24]        &   L[250]     &  104 &   0.833$\pm$0.091 &  0.931$\pm$0.029 &  0.121 &  0.924 \\
    L[24]        &   L[350]     &  104 &   0.223$\pm$0.106 &  0.932$\pm$0.034 &  0.141 &  0.898 \\
    L[24]        &   L[500]     &   61 &   0.332$\pm$0.150 &  0.853$\pm$0.048 &  0.145 &  0.872 \\
\hline
    L[70]        &   L[8(dust)] &  104 &   1.051$\pm$0.065 &  0.862$\pm$0.020 &  0.079 &  0.971 \\
    L[70]        &   L[24]      &  104 &  -0.015$\pm$0.044 &  0.927$\pm$0.013 &  0.057 &  0.987 \\
    L[70]        &   L[160]     &  104 &   1.153$\pm$0.063 &  0.886$\pm$0.019 &  0.085 &  0.963 \\
    L[70]        &   L[250]     &  104 &   0.777$\pm$0.079 &  0.867$\pm$0.024 &  0.107 &  0.941 \\
    L[70]        &   L[350]     &  104 &   0.224$\pm$0.100 &  0.862$\pm$0.031 &  0.134 &  0.912 \\
    L[70]        &   L[500]     &   61 &   0.418$\pm$0.146 &  0.781$\pm$0.045 &  0.149 &  0.885 \\
\hline
   L[160]        &   L[8(dust)] &  104 &  -0.225$\pm$0.064 &  0.989$\pm$0.020 &  0.067 &  0.974 \\
   L[160]        &   L[24]      &  104 &  -1.244$\pm$0.093 &  1.049$\pm$0.029 &  0.112 &  0.945 \\
   L[160]        &   L[70]      &  104 &  -1.243$\pm$0.079 &  1.124$\pm$0.024 &  0.096 &  0.963 \\
   L[160]        &   L[250]     &  104 &  -0.308$\pm$0.037 &  0.975$\pm$0.011 &  0.045 &  0.990 \\
   L[160]        &   L[350]     &  104 &  -0.923$\pm$0.080 &  0.976$\pm$0.025 &  0.093 &  0.958 \\
   L[160]        &   L[500]     &   61 &  -0.750$\pm$0.157 &  0.895$\pm$0.048 &  0.139 &  0.916 \\
\hline
   L[250]        &   L[8(dust)] &  104 &  -0.028$\pm$0.076 &  1.026$\pm$0.024 &  0.080 &  0.963 \\
   L[250]        &   L[24]      &  104 &  -0.893$\pm$0.106 &  1.074$\pm$0.034 &  0.130 &  0.924 \\
   L[250]        &   L[70]      &  104 &  -0.888$\pm$0.101 &  1.152$\pm$0.032 &  0.123 &  0.941 \\
   L[250]        &   L[160]     &  104 &   0.316$\pm$0.038 &  1.026$\pm$0.012 &  0.046 &  0.990 \\
   L[250]        &   L[350]     &  104 &  -0.634$\pm$0.044 &  1.003$\pm$0.014 &  0.050 &  0.985 \\
   L[250]        &   L[500]     &   61 &  -0.863$\pm$0.121 &  0.957$\pm$0.038 &  0.104 &  0.951 \\
\hline
   L[350]        &   L[8(dust)] &  104 &   0.272$\pm$0.097 &  1.061$\pm$0.032 &  0.118 &  0.938 \\
   L[350]        &   L[24]      &  104 &  -0.225$\pm$0.119 &  1.072$\pm$0.039 &  0.151 &  0.898 \\
   L[350]        &   L[70]      &  104 &  -0.228$\pm$0.125 &  1.156$\pm$0.041 &  0.155 &  0.912 \\
   L[350]        &   L[160]     &  104 &   0.946$\pm$0.079 &  1.024$\pm$0.026 &  0.096 &  0.958 \\
   L[350]        &   L[250]     &  104 &   0.632$\pm$0.042 &  0.997$\pm$0.014 &  0.050 &  0.985 \\
   L[350]        &   L[500]     &   61 &  -0.611$\pm$0.068 &  0.990$\pm$0.022 &  0.058 &  0.983 \\
\hline
   L[500]        &   L[8(dust)] &   61 &   0.508$\pm$0.178 &  1.114$\pm$0.060 &  0.155 &  0.898 \\
   L[500]        &   L[24]      &   61 &  -0.369$\pm$0.194 &  1.170$\pm$0.066 &  0.170 &  0.872 \\
   L[500]        &   L[70]      &   61 &  -0.535$\pm$0.216 &  1.281$\pm$0.073 &  0.191 &  0.885 \\
   L[500]        &   L[160]     &   61 &   0.838$\pm$0.177 &  1.117$\pm$0.060 &  0.156 &  0.916 \\
   L[500]        &   L[250]     &   61 &   0.902$\pm$0.124 &  1.045$\pm$0.042 &  0.109 &  0.951 \\
   L[500]        &   L[350]     &   61 &   0.617$\pm$0.066 &  1.010$\pm$0.023 &  0.058 &  0.983 \\
\hline
\end{tabular}\\
Col.(1)-(2): names of multi-$\lambda$ luminosities;
Col.(3): the number of sample galaxies (after excluding dwarf galaxies) used for the fitting procedures;
Col.(4)-(5): the coefficients $a$ and $b$ of the nonlinear fit: $\log_{10}(y)=$a$+$b$\log_{10}(x)$. All the luminosities are in units of L$_{\odot}$;
Col.(6): the standard deviation $s$ of the fitting residuals;
Col.(7): the coefficient $r$ of the Spearman Rank-order correlation analysis.
\end{table*}

Figure 2 shows the correlations among various monochromatic luminosities, including L[8(dust)], L[24], L[70], L[160], L[250], 
L[350] and L[500]. The panels on the diagonal from left up to right down are the correlations between the monochromatic luminosities 
and themselves. So the figure is symmetrical. The numbers of galaxies in every panels are listed in Table 1. Using the two-variable 
regression, we obtain the best nonlinear fits ($\log_{10}(y)=$a$+$b$\log_{10}(x)$). The best nonlinear fits are shown as the 
solid line in each panel of Figure 2, and the fitting parameters are listed in Table 1, too. Then the standard deviation of 
the fitting residuals and the coefficient of the Spearman Rank-order correlation analysis are also listed in Table 1.

From Figure 2, we can see the star-forming galaxies in our sample show good correlations among various monochromatic luminosities.
L[8(dust)] correlates most tightly with L[160] than with others. For L[24], the correlation with L[70] shows the largest Spearman 
Rank-order correlation analysis coefficient (0.987); L[70] also tend to correlate with L[24] most tightly. While the correlations 
between L[24] and other monochromatic luminosities, even L[8(dust)], the other nearest waveband to 24$\mu$m, are not as tight 
as the correlation between L[24] and L[70]. From L[70] to L[500], the coefficients of the Spearman Rank-order correlation 
analysis between them and L[24] decrease monotonically with increasing wavelength \citep{lam13}, while the fitting errors increase gradually. 

\section{Discussion}

\subsection{L[8(dust)] and Cool Dust Emission}

The monochromatic SPIRE luminosities, including L[250] \& L[350] \& L[500], even the $Spitzer$ L[160] are clearly dominated 
by cool dust emission. The L[8(dust)] correlates with monochromatic FIR luminosities more tightly, particularly with L[160], 
rather than with L[24] \& L[70], indicating that the heating sources for the 8$\mu$m(dust) and FIR emission may be same. For 
a local star-forming galaxy, its 160$\mu$m luminosity could be simply divided to two different parts: from HII regions, related 
with massive young hot stars; from photo-dissociation regions (PDRs) or much cooler regions excited by interstellar radiation 
field (ISRF), related with evolved stars. The dust emission related with ISRF has been found to be prominent for some 
very nearby galaxies' output at rest-frame 160$\mu$m \citep{bendo10,bendo12}. The rest-frame 8$\mu$m(dust) luminosity of the 
local star-forming galaxies could also be divided to two different parts: VSGs' continuum and spectral feathers from C-C stretch 
and C-H bending of ionized PAHs. The continuum of VSGs is connected with on-going star formation (see next subsection). Although 
the carriers of PAHs could be considered as some types of VSGs with smaller scale ($\sim$20\AA), the excitation for the PAHs' 
features and VSGs' continuum could be definitely different. The obvious decrease of PAH emission in low metallicity galaxies 
has been detected. And the hard or intensive radiation may be responsible for this decrease \citep{engelbracht05, hogg05, ohalloran06, madden06, wu07, engelbracht08, galametz09, khramtsova13}. 
Given the hypothesis is true, the ionized PAHs should be fragile. 

From these considerations, PAHs should tend to settle in the PDRs instead of HII regions, since the PDRs are dominated by 
ISRF and could shield the intense UV photons \citep{tielens08}. In order to check this hypothesis, it is necessary to search 
for the probable coincidence of the distributions of the PAHs and ISRF. The ISRF can be traced by middle-age or old stars. 
In this work, we use L[3.6], the rest-frame IRAC 3.6$\mu$m luminosities, to represent the ISRF. The rest-frame 3.6$\mu$m emissions 
can be used to compute the total stellar mass in galaxies without remarkable AGNs \citep{li07, cao08}, since the emission 
in this wavelength range is dominated by the photons from photosphere of red giants or supergiants, such as (post-) Asymptotic 
Giant Branch (AGB \& p-AGB) stars \citep{zhu10, wen13}. 

\begin{figure}
\includegraphics[scale=1]{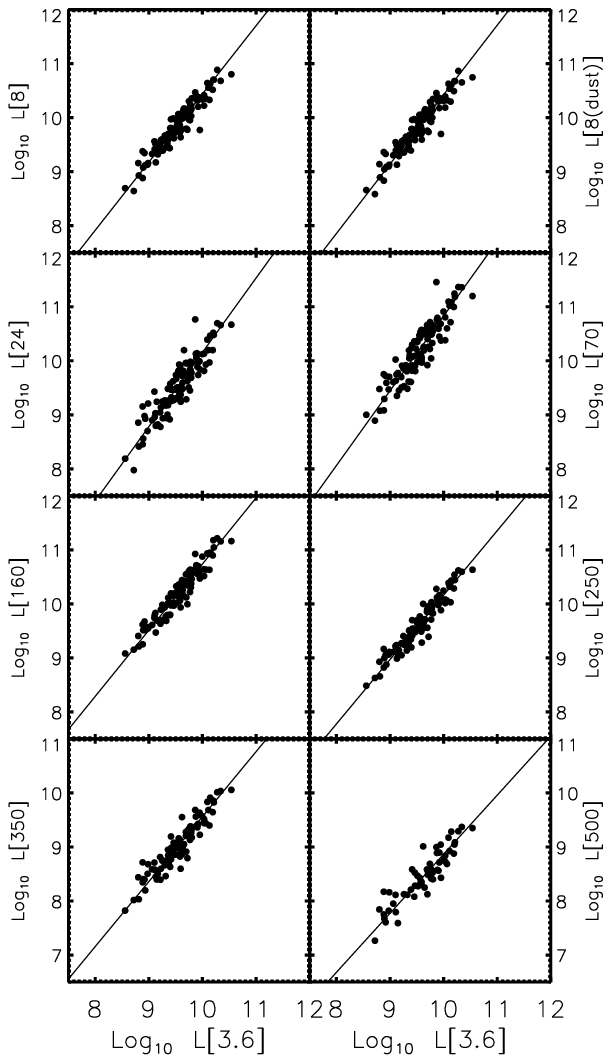}
\caption{Correlations between 3.6$\mu$m and the other monochromatic IR luminosities for star-forming galaxies.}
\end{figure}

\begin{table*}
\centering
\caption{correlation coefficients between 3.6$\mu$m and monochromatic IR luminosities}
\begin{tabular}{llrcccc}
\hline
{$x$} & {$y$} & {$N$}  & {$a$} & {$b$} & {$s$} & {$c$} \\
\hline
{(1)} & {(2)} & {(3)}  & {(4)} & {(5)} & {(6)} & {(7)} \\
\hline
   L[3.6]  & $vs$ $~~$ L[8]        &  104 &  -2.175$\pm$0.103 &  1.261$\pm$0.033 &  0.105 &  0.943 \\
   L[3.6]  & $vs$ $~~$ L[8(dust)]  &  104 &  -2.422$\pm$0.112 &  1.284$\pm$0.036 &  0.112 &  0.940 \\
   L[3.6]  & $vs$ $~~$ L[24]       &  104 &  -2.838$\pm$0.149 &  1.286$\pm$0.048 &  0.153 &  0.904 \\
   L[3.6]  & $vs$ $~~$ L[70]       &  104 &  -3.005$\pm$0.164 &  1.383$\pm$0.053 &  0.164 &  0.909 \\
   L[3.6]  & $vs$ $~~$ L[160]      &  104 &  -1.502$\pm$0.103 &  1.224$\pm$0.033 &  0.101 &  0.951 \\
   L[3.6]  & $vs$ $~~$ L[250]      &  104 &  -1.759$\pm$0.091 &  1.192$\pm$0.029 &  0.090 &  0.955 \\
   L[3.6]  & $vs$ $~~$ L[350]      &  104 &  -2.456$\pm$0.105 &  1.202$\pm$0.034 &  0.108 &  0.934 \\
   L[3.6]  & $vs$ $~~$ L[500]      &   61 &  -2.061$\pm$0.186 &  1.092$\pm$0.060 &  0.141 &  0.892 \\
\hline
\end{tabular}\\
The definition of columns are the same as in Table 1.
\end{table*}

The correlations between L[3.6] and other monochromatic luminosities are illustrated in Figure 3. L[8] \& L[8(dust)] represent 
the rest-frame 8$\mu$m luminosities without and with stellar continuum subtraction. The best nonlinear fits by using the two-variable 
regression are plotted as the solid lines in every panels, and the fitting parameters are listed in Table 2. The correlations 
of L[3.6] vs. L[250] and L[3.6] vs. L[160] are the tightest, then is the correlations of L[3.6] vs. L[8] and L[3.6] vs. L[8(dust)]. 
The cool dust continuum dominating L[160] and L[250], is unambiguously related with evolved stars. Therefore, the MIR features 
of C-C stretch and C-H bending of ionized PAHs seem to relate with the evolved stars as well, excited by moderate ISRF rather 
than by intensive radiation fields. This conclusion is as same as the previous findings \citep{haas02, boselli04, bendo06, bendo08, lu08, boquien11}. 
\citet{slater11} demonstrated the weeker PAH emission in some of active HII regions in Large Magellanic Cloud compared with those 
evolved ones, and some PAH emissions outside the galactic disks even have been detected in some nearby galaxies \citep{engelbracht06,irwin06,kaneda10}.
Resently, by using the Herschel's data, some researchers also discovered the 8$\mu$m emission is consistent with heating by 
the diffuse interstellar medium \citep{calapa14, ciesla14, lu14}.
Accordingly, compared with the other monochromatic MIR luminosities dominated by the VSGs' continuum instead of PAHs' features, 
L[8(dust)] seems not to be a good star formation tracer for star forming galaxies in local universe, particularly for the 
galaxies with weak star formation activities, whose fraction of emission from HII regions is significant small \citep{wu05, wu07, zhu08}.

\begin{figure}
\includegraphics[scale=1.0]{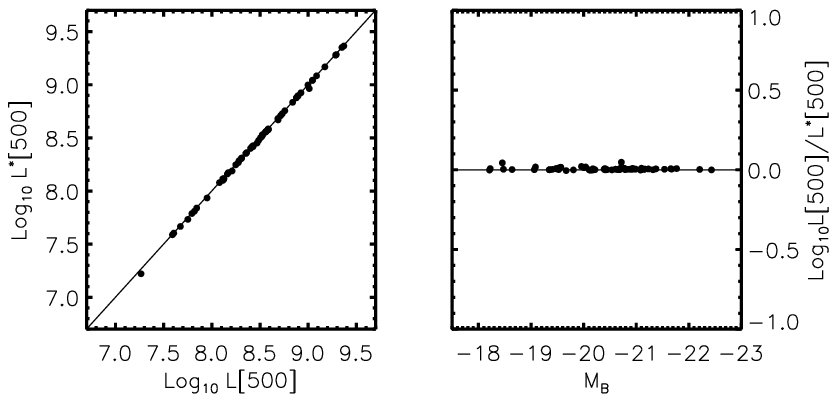}
\caption{Left: the correlation between L[500] (after K-correction) and the L$^{*}$[500] extrapolated from the other band observation 
luminosities (from 24 to 350$\mu$m); Right: the correlation between M$_B$ and L[500]/L$^{*}$[500]. }
\end{figure}

Additionally, L[350] shows good correlation with L[3.6]. But the correlation between L[500] and L[3.6] is the worst among all the correlations in Table 2. 
The sub-millimeter excess has also been found in some nearby galaxies based on the observations of $Herschel$ \citep{auld13}. 
The excess is usually accredited to the radiation from cold dust, whose temperature is lower than 10K \citep{skibba12}. But 
in this work, on the basis of available data, we can not determine the fraction of cold dust at 500$\mu$m emission. In Figure 4., 
we plot the correlation between L[500] (after K-correction) and the one extrapolated from the other band observation luminosities 
(from 24 to 350$\mu$m), and the right panel of Figure 4 is their ratio as the function of B-band absolute magnitude. We could 
see that, compared with the extrapolated 500$\mu$m luminosity, L[500] has not apparent enhancement, which leads to the conclution 
that the cold dust emission may be needless for the galaxies in our sample. This could also be due to the absence of dwarfs 
in our sample, while sub-millimeter excess has often be found in local low metallicity dwarf galaxies \citep{bolatto00, galliano05, galliano11}.
Furthermore, the thermal emission of cold dust is not the only possible explanation for the sub-millimeter excess \citep{dale12}. 
Others, such as spinning dust \citep{bot10}, could enhance the sub-millimeter emission as well. 

\subsection{L[24] and Warm Dust Emission}

Figure 2 shows that the correlation between L[24] vs. L[70] is tighter than the rests. Traditionally, L[70] is thought to 
be dominated by the warm dust continuum; while L[24] is considered from VSGs heated stochastically by single UV photon emitted 
by young hot stars. Therefore, it is necessary to check the correlations between monochromatic IR luminosities and the star 
formation rate (SFR).

Here, we use the H$\alpha$ luminosity as the star formation rate tracer \citep{kennicutt98}. The optical H$\alpha$ emissions 
suffer from the dust extinctions from both Milky Way and the host galaxy. The foreground Galactic extinction was first corrected 
by assuming the \citet{cardelli89}'s extinction curve and R$_{\rm V}$=3.1. The intrinsic extinction correction was performed, 
based on the color excess E(B-V) which was acquired from the Balmer decrement $F_{\rm H\alpha}/F_{\rm H\beta}$ \citep{calzetti01}.
The optical spectra of SDSS were taken with 3$\arcsec$ diameter fibers, thus the measured H$\alpha$ emission fluxes were only from
the central regions for most low redshift galaxies and the corresponding aperture correction is needed. We adopted the method by
\citet{hopkins03} and used their Equation (A2) to obtain the H$\alpha$ luminosity of the whole galaxy \citep{zhu08}. The aperture 
correction to H$\alpha$ luminosities was performed based on the assumption that the distributions of H$\alpha$ emission in 
galaxies are similar to those of the continuum (SDSS-$r$) emission. 

\begin{figure}
\includegraphics[scale=1.0]{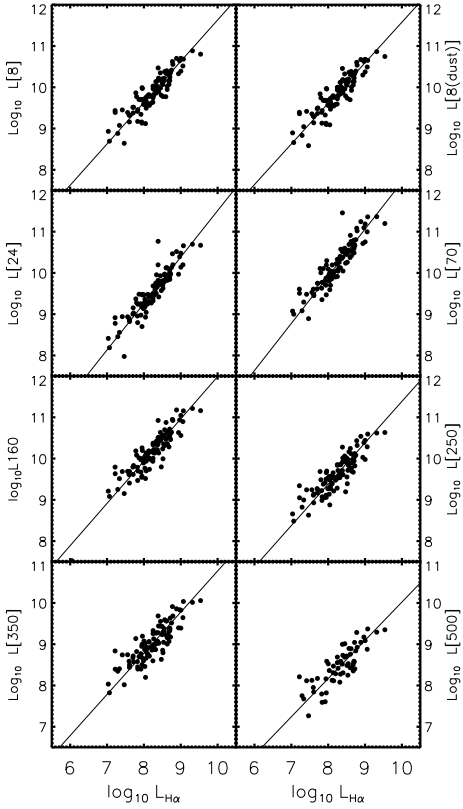}
\caption{Correlations between H$\alpha$ and monochromatic IR luminosities for star-forming galaxies.}
\end{figure}

\begin{table*}
\centering
\caption{correlation coefficients between H$\alpha$ and monochromatic IR luminosities}
\begin{tabular}{llrcccc}
\hline
{$x$} & {$y$} & {$N$}  & {$a$} & {$b$} & {$s$} & {$c$} \\
\hline
{(1)} & {(2)} & {(3)}  & {(4)} & {(5)} & {(6)} & {(7)} \\
\hline
    L[H$\alpha$]  & $vs$ $~~$ L[8]        & 104  & 1.617$\pm$0.122 & 1.001$\pm$0.042 & 0.151 & 0.909 \\
    L[H$\alpha$]  & $vs$ $~~$ L[8(dust)]  & 104  & 1.739$\pm$0.123 & 0.984$\pm$0.042 & 0.148 & 0.909 \\
    L[H$\alpha$]  & $vs$ $~~$ L[24]       & 104  & 0.789$\pm$0.100 & 1.049$\pm$0.034 & 0.129 & 0.936 \\
    L[H$\alpha$]  & $vs$ $~~$ L[70]       & 104  & 0.781$\pm$0.112 & 1.142$\pm$0.038 & 0.146 & 0.933 \\
    L[H$\alpha$]  & $vs$ $~~$ L[160]      & 104  & 1.745$\pm$0.123 & 1.024$\pm$0.042 & 0.161 & 0.898 \\
    L[H$\alpha$]  & $vs$ $~~$ L[250]      & 104  & 1.351$\pm$0.130 & 1.003$\pm$0.044 & 0.169 & 0.878 \\
    L[H$\alpha$]  & $vs$ $~~$ L[350]      & 104  & 0.802$\pm$0.144 & 0.997$\pm$0.049 & 0.187 & 0.850 \\
    L[H$\alpha$]  & $vs$ $~~$ L[500]      &  61  & 0.749$\pm$0.211 & 0.927$\pm$0.072 & 0.202 & 0.813 \\
\hline
\end{tabular}\\
The definition of columns are the same as in Table 1.
\end{table*}

Figure 5 shows the correlations between L[H$\alpha$] versus monochromatic IR luminosities. The best nonlinear fits by using 
the two-variable regression are plotted as the solid lines in Figure 5, and the fitting parameters are listed in Table 3. 
Apparently, L[24] \& L[70] correlate with L[H$\alpha$] most tightly in Figure 5. The scatter in the correlations between L[8(dust)] 
vs. L[H$\alpha$] is larger than the ones of L[24] vs. L[H$\alpha$] and L[70] vs. L[H$\alpha$]. From L[160] to L[500], the 
correlations with L[H$\alpha$] decrease gradually, similar as the tendencies with L[24] \& L[70] (see Section 3). Hence, compared 
with other monochromatic IR luminosities, L[24] \& L[70] should be considered as good SFR tracers, which is consistent with 
previous fingdings, such as \citet{calzetti05, wu05, calzetti07, kennicutt07, kennicutt09, calzetti10, kennicutt07, kennicutt09}.

\begin{figure}
\hspace{-1.2cm}
\includegraphics[scale=1]{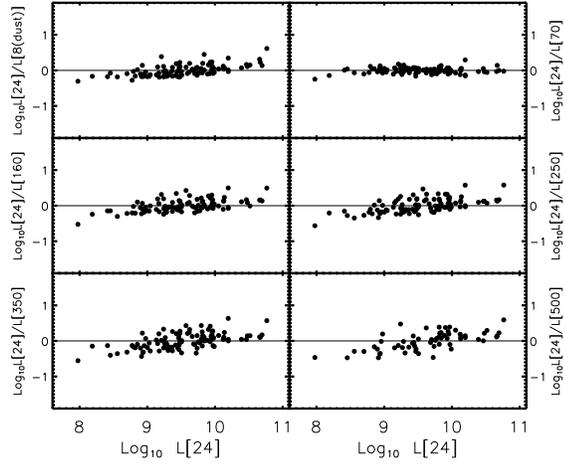}
\caption{Correlations between L[24] and the ratios of L[24] to other monochromatic luminosities. The ratios has been shifted 
to around zero by subtracting the statistic mean value in each panel. As the wavelength of the denominator increases, the mean 
value is 0.305, 0.668, 0.652, 0.090, -0.512, -1.176, respectively.}
\end{figure}

Now, we employ L[24] as the SFR tracer, and show the ratios between L[24] and the rest six monochromatic IR luminosities, 
as a function of L[24] in Figure 6. The ratios has been shifted to around zero by subtracting the statistic mean value in each panel. In Figure 6, we could see that except L[24]/L[70], the rest five ratios can not keep themselves as the approximate 
constants with increasing SFR. Therefore, compared with 24 and 70$\mu$m, the fraction of emissions related with young stars 
is smaller in other FIR wave bands, and even in 8$\mu$m(dust). Then, by using the luminosity ratio of 24 to 3.6$\mu$m as the 
tracer of specific star formation rate (SSFR), we demonstrate the correlations between SSFR and the ratios between L[24] and 
the rest six monochromatic luminosities in Figure 7. The panels in Figure 7 are similar as the corresponding ones in Figure 6, 
only the SFR is replaced by SSFR in x-axis and the scatter in each panels is smaller. It is clear in Figure 7, that the ratios 
between dust related with or without current star formations increase gradually with star formation activity, which indicates 
that the overestimation of SFR would be inevitable if we only calculate SFR only based on the FIR ($>$100$\mu$m) luminosity, 
particularly for the galaxies with significant emission from cool dust \citep{lam13}. 

\begin{figure}
\hspace{-1.2cm}
\includegraphics[scale=1.0]{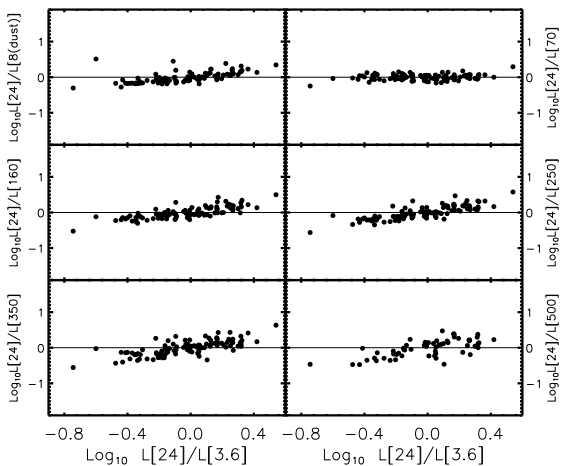}
\caption{Correlations between SSFR and the ratios of L[24] to other monochromatic luminosities. The SSFR was represented by 
the ratio of 24 to 3.6$\mu$m luminosities. Same as in Figure 6, the ratios has been shifted to around zero by subtracting the statistic mean value in each panel.}
\end{figure}

\subsection{Estimation of $L_{TIR}$ from L[8(dust)] \& L[24]}

We could suspect the deviation of $L_{TIR}$ calculation from one waveband flux, must be functioned by current star formation. 
This case has been discussed detailed by using KINGFISH \citep[Key Insights on Nearby Galaxies: a Far-Infrared Survey with $Herschel$;][]{kennicutt12} galaxies \citep{galametz13}. 
Here, we recalibrate the correlations between L[8(dust)] \& L[24] with $L_{TIR}$. $L_{TIR}$ was obtained from fitting six 
bands fluxes (except 8$\mu$m, because of the undefined contributions from PAHs) with a grey body emission and a MIR power-law 
continuum. The procedures were supplied by \citet{casey12} and \citet{casey13}. Only the sources with detectable fluxes larger 
than 3$\sigma$ at 500$\mu$m band were selected. The final sample includes $61$ galaxies.

\begin{figure}
\includegraphics[scale=1]{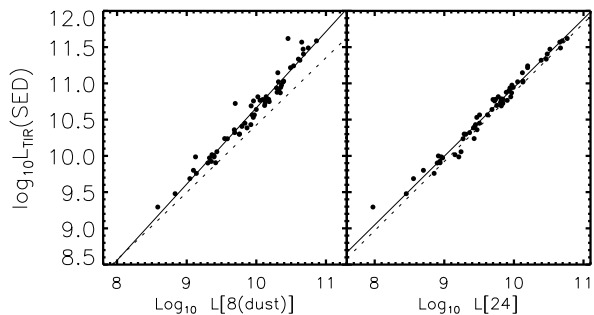}
\caption{Correlations between MIR luminosities and $L_{TIR}$ from SED fitting.The best nonlinear fits are illustrated as solid lines.
The dotted lines are the nonlinear fits in \citet{galametz13}.}
\end{figure}

\begin{table*}
\begin{center}
\caption{correlation coefficients between L[8(dust)] and L$_{TIR}$ from SED fitting}
\begin{tabular}{llrcccc}
\hline
{$x$} & {$y$} & {$N$}  & {$a$} & {$b$} & {$s$} & {$c$} \\
\hline
{(1)} & {(2)} & {(3)}  & {(4)} & {(5)} & {(6)} & {(7)} \\
\hline
 L[8(dust)]  & $vs$ $~~$ L$_{TIR}$ &  61  &   0.149$\pm$0.071 &  1.051$\pm$0.022 & 0.062  &  0.980 \\
 L[24]       & $vs$ $~~$ L$_{TIR}$ &  61  &   1.467$\pm$0.052 &  0.948$\pm$0.016 & 0.051  &  0.987 \\
\hline
\end{tabular}\\
The definition of columns are the same as in Table 1.
\end{center}
\end{table*}

L[8(dust)] \& L[24] are plotted against $L_{TIR}$ in Figure 8. The best nonlinear fits by using the two-variable regression 
are plotted as the solid lines. Table 4 lists the fitting parameters. The dotted lines in Figure 8 represent the fitting correlations 
for nearby galaxies from SINGS \& KINGFISH obtained by \citet{galametz13}. We find that the best nonlinear fit of L[24] vs. 
$L_{TIR}$ in this work is well consistent with the corresponding result in \citet{galametz13}. However, the slope of our nonlinear 
fit of L[8(dust)] vs. $L_{TIR}$ is a little steeper than that in \citet{galametz13}, and this discrepancy could be due to 
the different galaxy population between these two samples: the galaxies in \citet{galametz13} are all nearby ones, even include 
some optical dwarfs, but lack of IR luminous galaxies (locate at up-right in the left panel of Figure 8), 
which prefer to have higher SFRs and lower ratios of L[8(dust)] to $L_{TIR}$.

\begin{figure}
\includegraphics[scale=1]{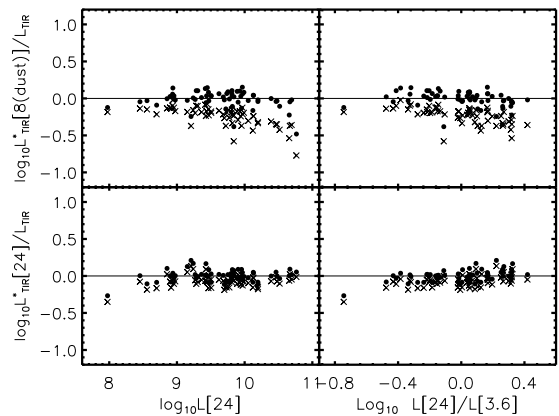}
\caption{Comparison between $L$$^{*}$$_{TIR}$ estimated by using L[8(dust)] \& L[24] and the $L_{TIR}$ from SED fitting, 
as a function of SFR \& SSFR, represented seperately by L[24] \& the luminosity ratio of 24 to 3.6$\mu$m. 
Solid circle in each panel means the $L$$^{*}$$_{TIR}$ estimated by using the formulae derived by ourselves (the parameters 
are demonstrated in Table 4); cross means the $L$$^{*}$$_{TIR}$ estimated by using the equations by \citet{galametz13}.}
\end{figure}

In order to check it, we demonstrate the $L_{TIR}$ ratios, between the one (represented as $L$$^{*}$$_{TIR}$ in this subsection) 
estimated by using monochromatic MIR luminosities (both L[8(dust)] \& L[24]) and the one from SED fitting, as a function of 
SFRs and SSFRs in Figure 9. Here, the SFR is represented by L[24] and the SSFR is represented by the luminosity ratio between 
at rest-frame 24 and 3.6$\mu$m. In Figure 9, solid circle means the $L$$^{*}$$_{TIR}$ estimated by using the nonlinear fitting 
parameters (Table 4); cross means the $L$$^{*}$$_{TIR}$ estimated by using the equations in \citet{galametz13}. In Figure 9, 
We could find that the $L$$^{*}$$_{TIR}$ calculated based on the formulae derived by us are larger than the $L$$^{*}$$_{TIR}$ 
calculated based on the formulae presented by \citet{galametz13}, particularly when utilizing the L[8(dust)] to estimate the 
$L$$^{*}$$_{TIR}$. Additionally, we could see, for the galaxies with higher SFRs or higher SSFRs, the ratio between the $L$$^{*}$$_{TIR}$ 
derived from L[8(dust)] (use both our and \citet{galametz13}'s formulae) and $L_{TIR}$ from SED fitting descends gradually, 
which indicates again that L[8(dust)] is not as good as L[24] and L[70] to trace on-going star formations. For the galaxies 
with higher SSFRs, the ratio between the $L$$^{*}$$_{TIR}$ derived from L[24] and $L_{TIR}$ from SED fitting increases slightly, 
but this tendency can not be seen in the left-bottom panel. One of the probable explanations is the galaxies with higher SSFR, 
rather than higher SFR, lack of cool dust grains, so the overestimation of $L$$^{*}$$_{TIR}$ derived from L[24] is unavoidable; 
another possible explanation is that the radiation field of the whole galaxies has been dominated by the young hot stars for 
the galaxies with higher SSFR. Consequently, for the galaxies with intense star formation, if we only employ MIR luminosity 
dominated by the emissions related with young stars to compute the total IR luminosity, the overestimation seems to be inevitable; 
whilst if we employ the formulae, which were derived on the basis of the monochromatic MIR luminosity influenced significantly 
by the spectral features of PAHs, the underestimation of the total IR luminosity would not be impossible. 

\section{Summary}

We present and analyze the correlations among L[8(dust)], L[24], L[70], L[160], L[250], L[350] \& L[500] for a sample of galaxies 
selected from the $Spitzer$ \& $Herschel$ data in two northern SWIRE fields. The galaxies in our sample represent the local 
star-forming galaxies, lack of optical and IR dwarfs, as well as ULIRGs. The main results described in this paper can be summarized as follows.

1. The L[24] \& L[70] of star-forming galaxies are found to be well correlated with each other, then correlated with 
extinction-corrected H$\alpha$ luminosities. These correlations indicate the similarity of heating sources of VSGs and warm 
dust. Therefore, estimating SFRs by using L[24] \& L[70] would be more available than using other monochromatic IR luminosities. 
Nevertheless, only employing the monochromatic IR luminosities that are primarily relevant with young hot stars may overestimate the $L_{TIR}$.

2. For FIR emissions longer than 100$\mu$m, the contributions from warm dust vanished gradually with increasing wavelength. 
Cool dust excited by ISRF controls the emissions in these FIR wave bands, especially for the galaxies with quiescent star formation. 
Hence, a color correction would be useful when obtaining the SFR by using FIR ($>$100$\mu$m) luminosities. 

3. L[8(dust)] is composed of continuum of VSGs and character spectral features of PAHs. L[8(dust)] is well correlate with 
FIR ($>$100$\mu$m) luminosities, especially with L[160] \& L[250]. So we suspect the C-C stretch and C-H bending of ionized 
PAHs are mainly excited by ISRF, and the features of PAHs follow the cool dust emission, rather than warm dust or VSGs' emission 
related with star formation. L[8(dust)] could be used to derive the SFR, but may be not as good as L[24] \& L[70]. We re-scale 
the correlations between the $L_{TIR}$ and L[8(dust)]. But, for galaxies with higher SFR or SSFR, underestimation of $L_{TIR}$ 
would be inevitable by only using L[8(dust)] or other monochromatic MIR luminosities dominated by the spectral features of PAHs.

\section*{Acknowledgments}

We thank Caitlin M. Casey and her collaborators for helping us revising their procedures for fitting IR SED, and we thank the anonymous referee for constructive comments and suggestions.
We thank Dr Z.-M. Zhou for smoothing the lauguage.
We acknowledge Drs. K. Xu, N.-Y. Lu, Z.-M. Zhou and Mam I Lam for advice and helpful discussions.

This project is supported by the National Natural Science Foundation of China (Grant number 11173030), the Ministry of Science and Technology of the People's republic of China through grant 2012CB821800 and 2014CB845705, and by NSFC grants 11225316, 11078017 and 10833006. 
This work is also supported by the Strategic Priority Research Program "The Emergence of Cosmological Structures" of the Chinese Academy of Sciences, Grant No. XDB09000000.”

This work is based on observations made with the $Spitzer$ Space Telescope, which is operated by Jet Propulsion Laboratory of the California Institute of Technology under NASA Contract 1407. 
This research has made use of the NASA/IPAC Infrared Science Archive, which is operated by the Jet Propulsion Laboratory, California Institute of Technology, under contract with the National Aeronautics and Space Administration.
We thank the useful SWIRE catalogs from NASA/IPAC Infrared Science Archive. 

Herschel is an ESA space observatory with science instruments provided by European-led Principal Investigator consortia and with important participation from NASA.
This research has made use of data from HerMES project (http://hermes.sussex.ac.uk/). HerMES is a Herschel Key Programme utilising Guaranteed Time from the SPIRE instrument team, ESAC scientists and a mission scientist. HerMES has be described in Oliver et al. 2012, MNRAS.
The HerMES data was accessed through the HeDaM database (http://hedam.oamp.fr) operated by CeSAM and hosted by the Laboratoire d'Astrophysique de Marseille.

We thank the useful SDSS database and the MPA/JHU catalogs. 
Funding for the SDSS and SDSS-II has been provided by the Alfred P. Sloan Foundation, the Participating Institutions, 
the National Science Foundation, the U.S. Department of Energy, the National Aeronautics and Space Administration, the Japanese 
Monbukagakusho, the Max Planck Society, and the Higher Education Funding Council for England. The SDSS Web Site is http://www.sdss.org/. 
The SDSS is managed by the Astrophysical Research Consortium for the Participating Institutions. The Participating Institutions are the 
American Museum of Natural History, Astrophysical Institute Potsdam, University of Basel, University of Cambridge, Case Western 
Reserve University, University of Chicago, Drexel University, Fermilab, the Institute for Advanced Study, the Japan Participation 
Group, Johns Hopkins University, the Joint Institute for Nuclear Astrophysics, the Kavli Institute for Particle Astrophysics and 
Cosmology, the Korean Scientist Group, the Chinese Academy of Sciences (LAMOST), Los Alamos National Laboratory, the Max-Planck-Institute 
for Astronomy (MPIA), the Max-Planck-Institute for Astrophysics (MPA), New Mexico State University, Ohio State University, University 
of Pittsburgh, University of Portsmouth, Princeton University, the United States Naval Observatory, and the University of Washington.  

\clearpage


\bsp

\label{lastpage}

\end{document}